\documentclass[prl,twocolumn,floatfix,superscriptaddress,showpacs,amsmath,amssymb]{revtex4}
\usepackage{bm}
\usepackage{graphicx}
\usepackage{dcolumn}
\usepackage{amsmath}
\usepackage{amssymb}
\usepackage{epsfig}

\begin{document}
\title{Comment on ``Solving the mystery of booming sand dunes"}
\author{Bruno Andreotti}
\author{Lenaic Bonneau}
\author{Eric Cl\'{e}ment}
\affiliation{PMMH, UMR7636 (CNRS), ESPCI Univ.~P6-P7, 10 Rue Vauquelin, 75005 Paris, France. }

\date{\today}

\begin{abstract}
We show here that the standard physical model used by \textit{Vriend et al.} to analyse seismograph data, namely a non-dispersive bulk propagation, does not apply to the surface layer of sand dunes. According to several experimental, theoretical and field results, the only possible propagation of sound waves in a dry sand bed under gravity is through an infinite, yet discrete, number of dispersive surface modes. Besides, we present a series of evidences, most of which have already been published in the literature, that the frequency of booming avalanches is not controlled by any resonance as argued in this article. In particular, plotting the data provided by \textit{Vriend et al.} as a table, it turns out that they do not present any correlation between the booming frequency and their estimate of the resonant frequency.
\end{abstract}

\maketitle
\section{Acoustics in sand dunes}
It is a well-known fact that sand dunes present a layered structure due to successive avalanches and capillary water retention. We focus here on the acoustic propagation within the metre scale surface layer composed of dry sand. According to \textit{Vriend et al.} [2007], in [\textit{Andreotti} 2004; \textit{Bonneau et al.} 2007], we ``only consider low-speed surface waves of around 50 m/s'' while ``booming results from the propagation of body waves, not surface waves''. This statement reflects a misunderstanding of acoustic waves in granular media. In fact, we have shown that, contrarily to ordinary elastic solids, bulk modes (the `body waves') do not exist at all [\textit{Bonneau et al.} 2007]. Instead, there is an infinite yet discrete number of surface modes (not a single one) with a dispersive propagation (i.e. not a single wave speed independent of vibration frequencies). This has been directly evidenced in the field and in a lab experiment (Figure~1), and reproduced by an independent team [\textit{Jacob et al.} 2007].

The explanation is simple. Sand is a divided medium that presents non-linearities of geometrical origin. As the grains do not have plane/plane contact, they behave like a spring whose stiffness depends on the normal force applied to put them into contact. As a consequence, the speed of sound $c$ depends on confining pressure $P$ --~roughly as $c \propto Z^{1/3} P^{1/6}$, where $Z$ is the effective number of contacts per grain. A striking consequence is that the order of magnitude of the propagation speed $c$ under a pressure of $\sim 1~m$ of sand is lower than the sound velocity in air (although density is $10^3$ times larger). Therefore the effective compressibility is extraordinarily small! This property cannot be explained without involving geometrical effects at the scale of a grain. This dependence on pressure has been proven several times experimentally and numerically for moderately large $P$ ($\sim 10-100$~m of sand in \textit{Jia et al.}). Due to gravity-induced pressure ($P \propto \rho gz$), the surface layer of a dune does not constitute a homogeneous system as $c$ increases with depth. Thus, no plane wave Fourier mode can exist in such a medium; only an infinite number of surface modes guided by the sound speed gradient may propagate. Consequently, a mode labelled $n$ is localized within a depth $n\,\lambda$ below the surface ($\lambda$ is the wavelength) i.e. in a zone where the typical pressure is $P\sim\rho g n \lambda$. Thus, the surface propagation velocity increases with the mode number $n$ in a similar way as $c(P)$ (between $n^{1/4}$ and $n^{1/3}$ typically). So, even in the limit of an infinite depth (no finite-depth layering), gravity produces a wave-guide effect, but no resonance.

Considering now a layer of sand of finite depth $H$, a second wave-guide effect get superimposed to the first one and resonant modes may appear. As the system is still not homogeneous due to gravity, these resonant modes are not Fourier modes. By definition, they do not propagate --~i.e. they have a vanishing group velocity~-- so that they correspond to cut-off frequencies of the system: no wave can propagate at a velocity smaller than the first resonant frequency $f_R$. The influence of the finite depth $H$ is in fact limited to a very narrow range of frequencies. As soon as the depth $H$ is larger than the wavelength (in practice, for a frequency $f$ $25\%$ above $f_R$), it can be considered as infinite and the gravity effect prevails [\textit{Bonneau et al.} 2007].

Then, it is easy to realize that if one strikes such a gradient-index medium with a sledge hammer (basically, the procedure used by \textit{Vriend et al.}), a series of wave packets corresponding to the different surface modes will be propagated that are related to the gravity induced index gradient and not to the effect of the finite depth $H$. Moreover, this procedure only gives access to the group velocity at the mean frequency of excitation --~yet not controlled nor specified by \textit{Vriend et al.}~-- and not at all at the frequency $f$ of spontaneous booming. Consequently, from such a procedure, just by reading the multi-modal structure of a seismograph obtained using \textit{only surface} transducers, one cannot conclude to the existence of multiple layers nor determine the resonant frequencies. Actually, if \textit{Vriend et al.} were right, they could easily provide a crucial test by performing measurements with transducers buried in the bulk, as those reported in \textit{Andreotti} [2004] (see Auxiliary figures). The vibration in the soil during booming avalanches is already strongly reduced at $60~$mm below the surface (Auxiliary Figure), which evidences directly that the song of dunes is not related to a resonant mode at the meter scale but to surface propagation. 

As a conclusion, the quantity determined by \textit{Vriend et al.} has probably nothing to do with the resonant frequency $f_R$. Besides, one may wonder why they did not measured directly the resonant frequencies $f_{R}$ for the seek of comparison with the booming frequency $f$. Using two different methods, we have ourselves performed such measurements using two different methods (Auxiliary Material). For a depth $H$ of dry sand of the order of $50~$cm, we find typically $70$~Hz; for $1$~m, the resonance becomes hardly visible and is below $50$~Hz. This is much smaller than the values found by \textit{Vriend et al.} and than the booming frequency $f$ ($100$~Hz in Morocco). So, as already evidenced in our previous papers, the phenomenon is not driven by a resonance effect. As shown in \textit{Bonneau et al.} [2007], the resonant frequency plays another role. As the surface waves do not propagate below $f_R$, the later controls the threshold for the booming phenomenon. A large enough layer of dry sand is required for the surface waves to propagate and thus for booming to occur. The best situation is thus an infinite layer of dry sand and not a layered one.

\section{Field evidences against the selection of the frequency by a resonance}
\textit{Vriend et al.} have missed another very important step of the argument presented in \textit{Andreotti} [2004]. The booming frequency $f$ is controlled by the shear rate $\dot \gamma$ (i.e. by the rate at which the grains collide with each other) in the shear zone separating the avalanche from the static part of the dune. Of course, $\dot \gamma$ depends on the way the avalanche is forced. When a granular flow is driven by a pressure gradient, e.g. by pushing with the hand or the bottom, $\dot \gamma$ can be varied and so does the emission frequency $f$. Besides, the reproduction of the phenomenon at small scale [\textit{Haff} 1986; \textit{Douady et al.} 2006; \textit{Bonneau et al.} 2007] is a direct proof that the acoustic emission is not related to the dune itself. Note in particular that controlled lab experiments have allowed to produce sustained booming (not short squeaks), varying continuously from gravity to pressure gradient driving.

Spontaneous avalanches on the slip face of dunes are driven by gravity in an homogeneous and steady way. In that case, \textit{only}, $\dot \gamma$ and thus $f$ are expected to scale as $\sqrt{g/d}$, with subdominant dependencies on the nature of the grains and the presence of cohesion. These conditions can be reproduced by a man-made slide only if the velocity of the body is constant. This is never the case in the movies provided by \textit{Vriend et al.} to illustrate their measurements. In the auxiliary animation, we show that a pulsed driving leads to a low frequency squeak similar to that measured by \textit{Vriend et al.}  (e.g. point at $70~$Hz in their figure 3). It is pretty clear that such an inhomogeneous and unsteady driving do not resemble the spontaneous avalanching process and thus, the avalanche has no reason to yield a constant booming frequency $f$. By contrast, during our $13$ field trips in Morocco from 2002 up to now, we have recorded around $20$ spontaneous avalanches and $100$ man-triggered slides, on barkhan dunes ranging from $4$ to $42$~m in height. This was done in different places and by different weather conditions. We always measured the same frequency $f$ within a tone. In particular, during our field trip of April 2007, we have performed a series of experiments for different flow thicknesses $H$, measured coherently by two independent techniques (see Auxiliary figures). Again, provided that avalanches were homogeneous and steady, the frequency $f$ turned out to be constant. This demonstrates that the frequency is not either selected by a resonance effect over the depth of the avalanching flow.

Figure~2 compares our predictions for the frequency $f$ of homogeneous booming avalanches ($f \sim 0.4\sqrt{g/d}$) and that proposed by the Caltech team ($f=f_R$). Once the data obtained by \textit{Vriend et al.} in Dumont are plotted (fig.~2b) --~and not anymore presented in a Table~-- no correlation can be observed between $f$ and what they claim to be the resonant frequency $f_{R}$ (see above). By contrast, with a choice of representation different from Vriend et al. (see tables and fig.~4 in Auxiliary Material), the relation between $f$ and $\sqrt{g/d}$ appears beyond any doubt (fig.~2a and solid line in  fig.~2b). This accumulation of evidences shows that the emission frequency $f$ is neither controlled by a resonance effect at the scale the avalanche as suggested by \textit{Douady et al.} [2006] nor at the scale of a superficial layer as suggested by \textit{Vriend et al} [2007]. 

The instability mechanism proposed in \textit{Andreotti} [2004] is based on the interaction between the plastic deformations (shear band) and the elastic compression (acoustics) of the granular material. The collisions between grains induce a transfer of momentum from translation to vibration modes. In turn, the surface waves tend to synchronise the collisions. Importantly, this phase synchronisation linear instability do not depend on the fact that the vibrations modes are propagating or resonant. Although some `mysteries' about singing dunes still remain to be solved, our explanation remains the single one consistent with all the existing observations.

\section{Auxiliary Material}
Auxiliary material for this comment gives the details necessary to understand our arguments and reproduce our measurements. There are three points of disagreement with Vriend et al. First, we show with a movie (Animation S1) that a man-made slide produces a booming similar to spontaneous avalanches only if the body is kept rigid and the sliding velocity as constant as possible. This movie is to be compared to those provided by Vriend et al. as Auxiliary Material, where the avalanches are systematically pulsed and inhomogeneous. Second, we show that the representation used by Vriend et al. to compare the frequency and the grain diameter (Fig. 4) does not allow to see the obvious correlation of Fig. 2a. We provide the table of the frequencies used in the figures of the comment (Fig. 2a, 2b and 4). A movie (Animation S1), a schematic of the measurement set-up (Fig. 5), granulometric measurements (Fig. 3) and different frequency spectra (Fig. 6) are provided to document the data points obtained in the Atlantic Sahara. It may serve as a reference to start a data-base of steady homogeneous booming avalanches. Third, we show that the method used by Vriend et al. to measure the first resonant frequency is not correct. We provide the measurement set-up for our controlled acoustic field experiments (Fig. 7a), including two methods to determine directly the resonant frequency (Fig. 8). Using these methods, we obtain a resonant frequency lower and not equal to the booming frequency.
\begin{figure}
\noindent\includegraphics{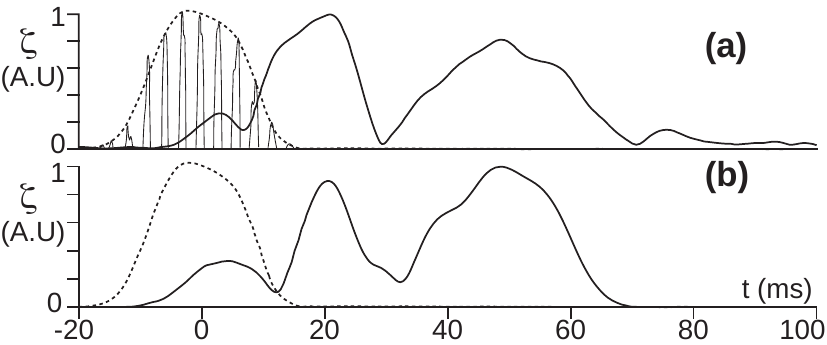} 
\caption{Field experiment: propagation of a synthetic gaussian wave-packet at $350~Hz$ at the surface of a $50~cm$ deep flat booming sand bed (proto-dune on a limestone plateau) i.e. without any layering effect. (a) Signal envelope $\zeta(t)$ measured by correlation with the emitted signal at $5~$cm (dotted line), $150~$cm (solid line)  from the source. (b) Comparison with the model proposed \textit{Bonneau et al. 2007}. The three wave-packets received are direct evidences of the existence of multiple propagative surface modes.}
\end{figure}
\begin{figure}
\noindent\includegraphics{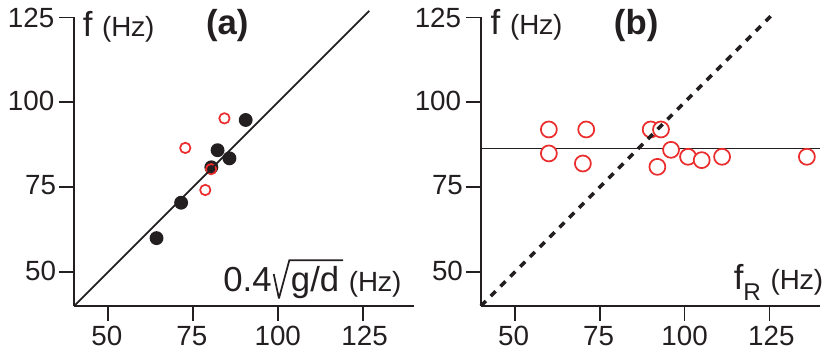} 
\caption{(a) Booming frequency $f$ as a function of grain size $d$. The plain symbols ($\bullet$)corresponds to homogeneous and steady avalanches (spontaneous or not) and the open ones ($\circ$) to the data of \textit{Vriend et al.} (see details in the Auxiliary Tables). (b) Plot of the data-set presented by \textit{Vriend et al.} as a Table (Large Dumont dune). Frequency $f$ of man-induced avalanches as a function of the (biased) resonant frequency $f_{R}$ (see text). On both plots a) and b), the solid line corresponds to our prediction. The dotted line is the prediction of {\it Vriend et al.} [2007]}
\end{figure}
\begin{table}
\begin{center}
\begin{tabular}{|c|c|c|}
\hline
 Place  & f (Hz) & d ($\mu$ m)  \\
\hline
Kelso  & $92.8 $& $200$\\
Sidi-Aghfenir & $103 $ & $163 $\\
Copiapo& $87$ & $210$\\
El Cerro Bramador & $75$&$270$\\
Sand moutain& $63$ & $340$\\
Tarfaya&$90$&$183$\\
\hline
\end{tabular}
\vspace*{0.3cm}
\end{center}
\caption{Data points for which we do know that (i) avalanches where spontaneous or at least homogeneous and steady (ii) the grain diameter was determined from samples taken in the middle of the slip face, at the place where the avalanche was recorded. When possible, the frequency was averaged over several realisations. \label{tab_compare}}
\end{table}
\begin{table}
\begin{center}
\begin{tabular}{|c|c|c|}
\hline
 Place  & f (Hz)\\
\hline
Big dune  & $93.5$\\
Eureka & $79.3$\\
Kelso& $103.6$\\
Dumont& $86.4 $\\
\hline
\end{tabular}
\vspace*{0.3cm}
\end{center}
\caption{Average frequency $f$ of the data points measured from Vriend et al.  No description of the conditions of experiment has been provided by these authors. The six movies they provided show the same technique to create avalanches, which are pulsed and inhomogeneous. No description of the place where the grains were sampled has been provided by these authors. The data obtained in Dumont on December 2005 have not been taken into account for several reasons: no recording is provided, no movie is provided, the width of the frequency peak is extremely large ($40\%$ of the central frequency) which is the signature of a small squeak induced by an unsteady avalanche (see our Auxiliary movie). So it cannot be used in a plot aiming to show data obtained in homogeneous and steady conditions.
\label{tab_compare}}
\end{table}
\begin{figure}
\noindent\includegraphics{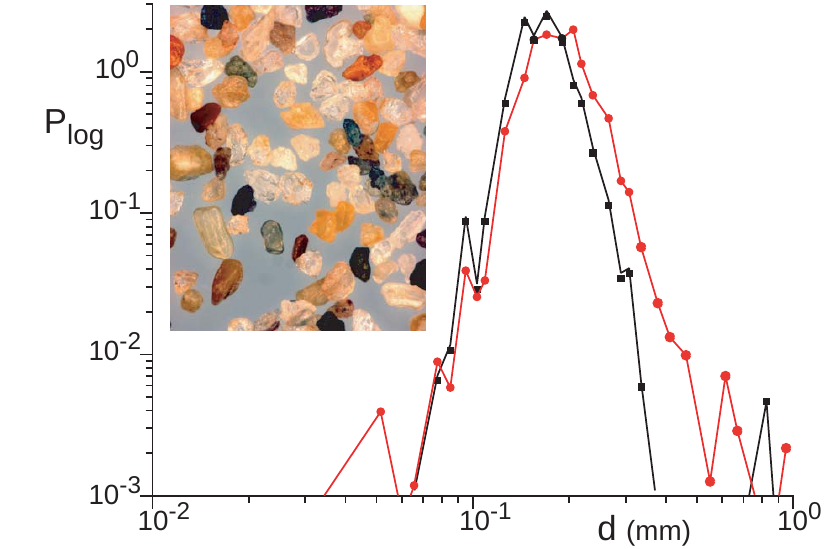} 
\caption{Probability density function (PDF), weighted in mass, of the logarithm of the grain size $d$. Red dots: $4~$m high barchan dune in Tarfaya  that booms at the frequency $f=90~$Hz (see Auxiliary film). Black squares: $42÷$m high mega-barkhan in Sidi Aghfinir  that boomed at a mean frequency $f=103~$Hz over 5 years (see Auxiliary film). The best fit by a log-normal distribution (a parabola in that representation) gives the centre of the distribution: $d=183\pm2~\mu$m and $d=163\pm2~\mu$m respectively. Inset: Photograph of few grains from the small barchan dune in Tarfaya.}
\end{figure}
\begin{figure}
\noindent\includegraphics{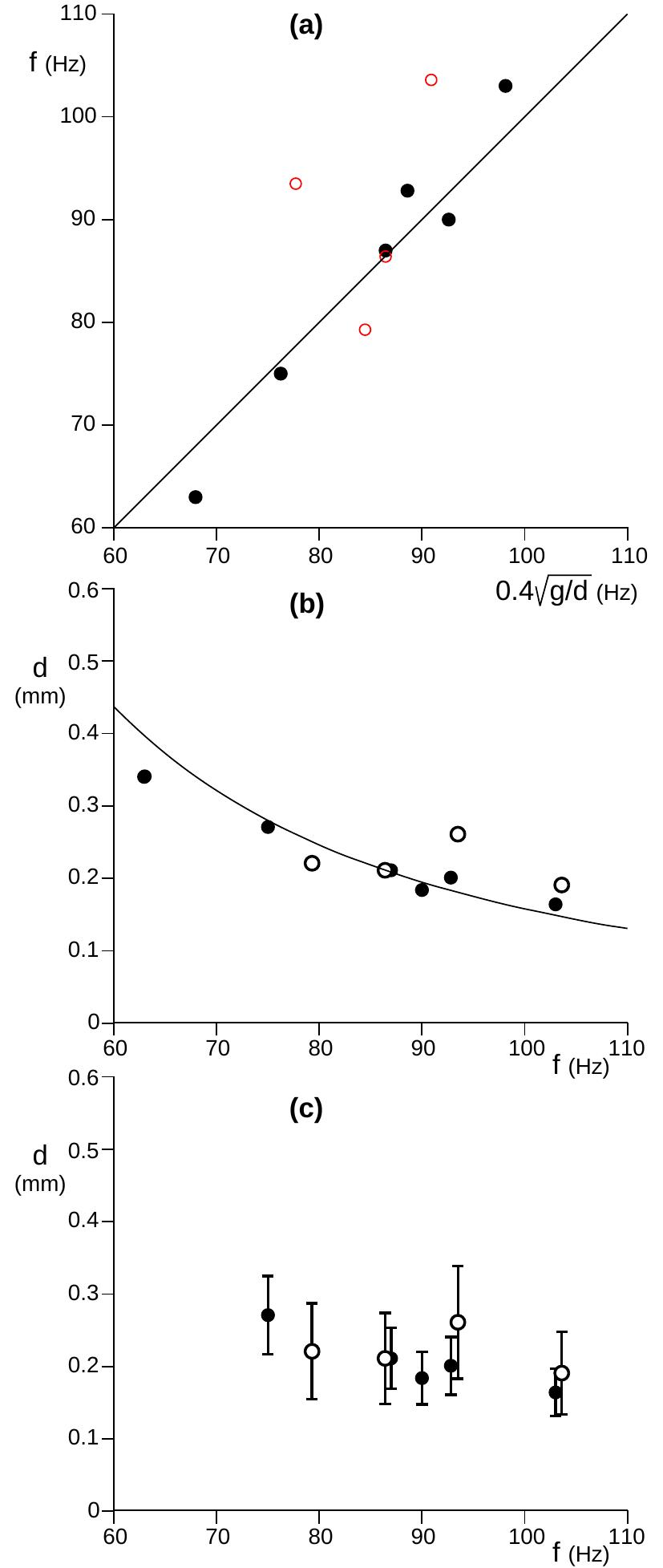} 
\caption{How can the same data show a clear scaling of the booming frequency $f$ with $\sqrt{g/d}$ and, as stated by Vriend et al, no correlation? (a) Replot of  figure 2a. The prediction corresponds to a straight line at $45^\circ$ in the graph (equal range of axis). The symbols show the conditions of booming: homogeneous and steady with a grain diameter $d$ determined in the middle of the avalanche slip face ($\bullet$) or pulsed and an unknown position at which sand was sampled ($\Large \circ$). (b) Same but plotting $d$ as a function of $f$. A large axis range is chosen for $d$ ($0$ to $600~\mu$m while the  dune grain size ranges from $150~\mu$m to $350~\mu$m), but the same narrow axis range for $f$. The solid line shows the prediction. (c) Same but adding error bars to indicate the width of the grain size distribution but letting no error bars on $f$, removing the last point on the left (Sand mountain), which significantly contributes to extend the dynamics, and removing the prediction. One obtains the plot presented by Vriend et al., which indeed shows no apparent correlation between $d$ and $f$.}
\end{figure}
\begin{figure}
\noindent\includegraphics{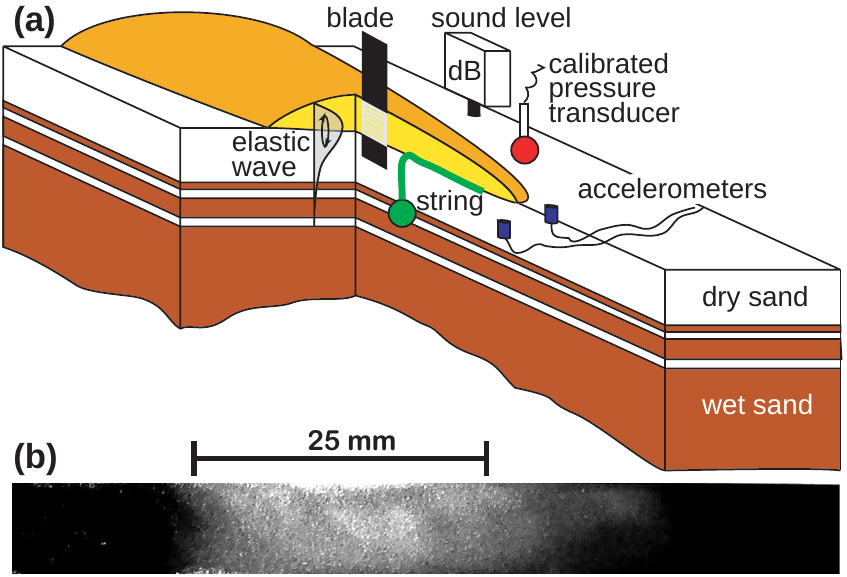} 
\caption{(a) Schematic of the measurement set-up used to characterise booming avalanches. The pressure in the air is measured using a calibrated pressure transducer at $10~$cm above the soil. A sound level apparatus using another microphone is filmed by the CCD camera (see Animation S1). The vibration of the grains is measured with piezo-electric accelerometers that can be put at different depth. The three measurements of the booming amplitude are checked to be consistent with the loud-speaker model (Andreotti 2004). The flowing depth is measured with two independent techniques. First, we bury a piece of string (green on the schematic) of diameter $4~$mm that can be curved with a very small force, attached at one hand to a large heavy object. The avalanching grains entrain the string and align it along the steepest slope, at the surface of the static part of the dune. Second, we prepare in advance a large number of metallic blades ($10~$mm by $150~$mm by $0.1~$mm) covered with black soot. The blade is then eroded during a fraction of the avalanche time (typically $1$ minute). Taking a photograph of the image of an homogeneous light, reflected by the light, the erosion intensity can be determined. As shown in controlled lab experiments, the grayscale is a function of the relative velocity between the flow and the blade. (b)  Picture of the blade after a $H=45~$mm thick avalanche. The left limit between black and white zones correspond to that determined by the string method. The grayscale profile is consistent with a plug flow separated from the static part of the dune by a shear band.}
\end{figure}	
\begin{figure}
\noindent\includegraphics{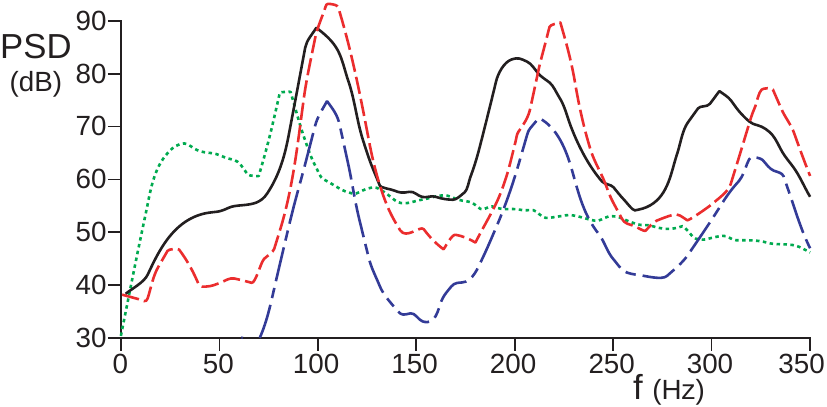} 
\caption{Power spectrum density (PSD) measured during homogeneous steady avalanches similar to spontaneous ones. The power radiated is the integral of the PSD. Dotted line: avalanche on a $4~$m high dune in Tarfaya ($d\simeq183~\mu$m). Dashed line: $H=65~$mm thick avalanche on a $42~$m high dune in Sidi-Aghfinir ($d\simeq163~\mu$m). Solid line: $H=15~$mm avalanche on the same $42~$m high dune. Dotted-dashed line: same avalanche  ($H=15~$mm), the signal coming from an accelerometer at $60~$mm below the surface.}
\end{figure}
\begin{figure}
\noindent\includegraphics{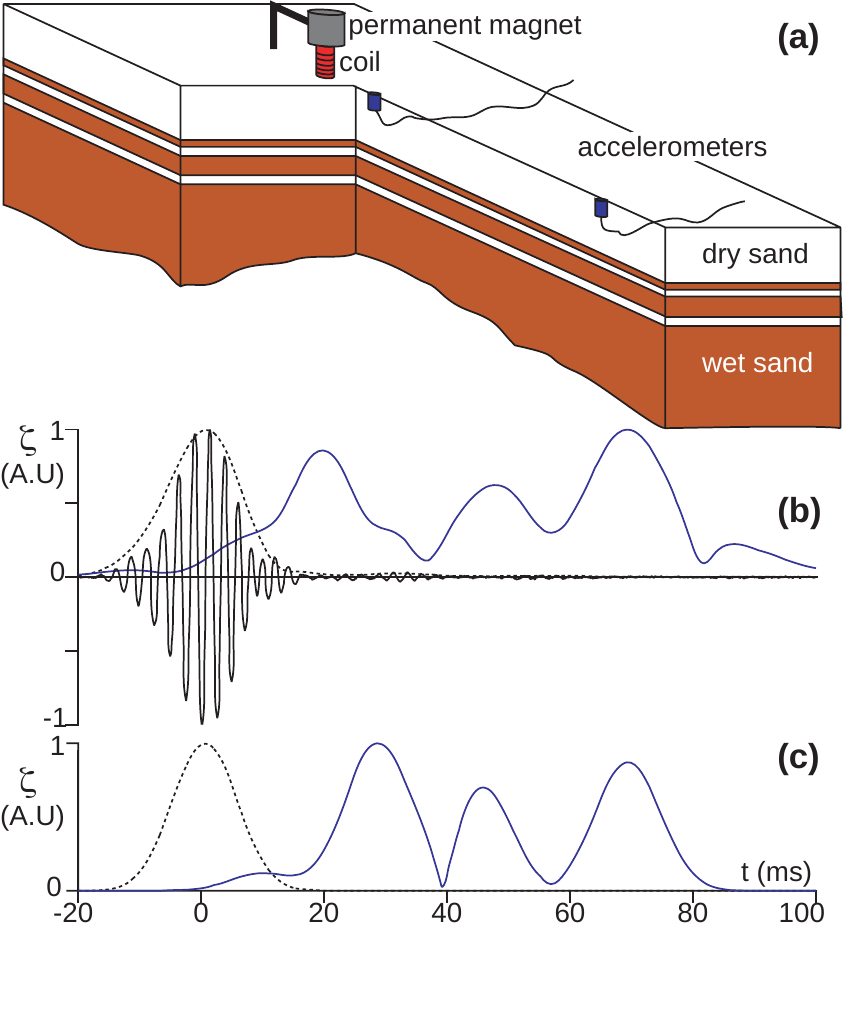} 
\caption{(a) Schematic of the measurement set-up used to characterise surface wave dispersion relation and in particular, the cut-off frequency (first resonant mode). The emitter is excited with an amplified sinusoidal source. It is composed of a coil glued around a sand-paper cylinder buried in the sand and submitted to the action of a fixed permanent loud-speaker magnet. This way, the force acting on the sand paper is imposed and, as inertia is small, transmitted to the grains. The transducers are piezoelectric accelerometers of diameter $10~$mm. This size is chosen to have a hierarchy of well-separated length scale: the wavelength (typically $40~$cm) is much larger than the transducer size, which is much larger than the grain size $d$. (b-c) Same as fig.~1 of this comment, but on the slip face of a booming dune. Field experiment: propagation of a gaussian wave-packet at $400~Hz$ along the slip face of a $42~$m high mega-barchan. (b) Signal envelope $\zeta(t)$ measured by correlation with the emitted signal at $10~$cm (dotted line), $220~$cm (solid line)  from the source. (b) Comparison with the model proposed \textit{Bonneau et al. 2007}. The multiple wave-packets received are direct evidences of the existence of multiple surface modes guided by the gravity induced index gradient.}
\end{figure}	
\begin{figure}
\noindent\includegraphics{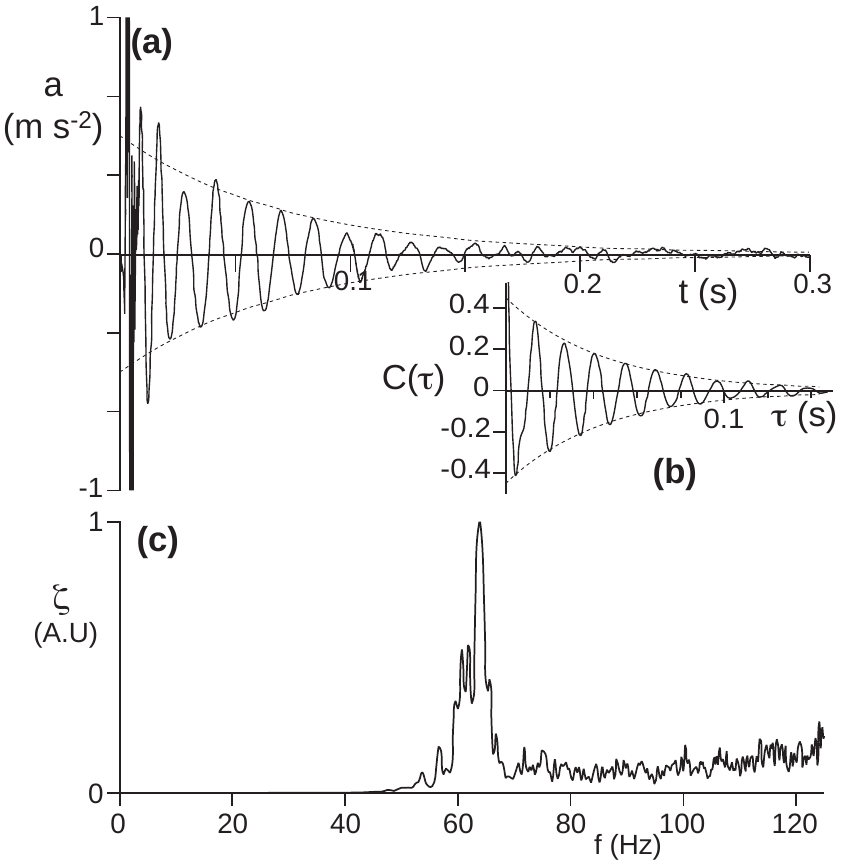} 
\caption{(a) The Makhnovist drum experiment: response of the booming dune to a normal tap containing constituting a broadband excitation. After the propagative modes excited have left, the resonant mode i.e. which does not propagate stays. The tail following the tap contains a well defined frequency that can be interpreted as the first compression resonant mode.  (b) Auto-correlation function of the signal shown in (a). The resonant frequency is around $73$~Hz for these particular conditions of field measurements.  The wet sand layer is at a depth around $H=50$~cm below the surface. [after Bonneau et al. 2006] (c) Amplitude $\zeta$ of vibration at $5~$m  from the source as a function of the frequency $f$, for permanent sinusoidal signals emitted at the surface of a 4m high booming dune. The wet sand layer is at a depth $H=50$~cm below the surface.}
\end{figure}	
\begin{figure}
\caption{\texttt{http://www.pmmh.espci.fr/fr/morphodynamique/Booming.mov} Movie of homogeneous and inhomogeneous booming avalanches in the Atlantic Sahara. From 0" to 38": homogeneous steady avalanche on a 42 m high barchan (d=163 microns). The sound level in dB can be read directly. The graduated stick gives the height of the free surface.  From 38" to 56": homogeneous steady avalanche on a 4 m high barchan (d=183 microns). From 56" to 70": homogeneous steady avalanche on the 42 m high barchan, recorded by the slider. The frequency is that of spontaneous events as the avalanche is driven by gravity. From 70" to 88": inhomogeneous unsteady avalanche produced by a pulsed slide. The frequency of the squeaks is not that of spontaneous events as the avalanche is driven by the pressure exerted by the bottom of the slider. This movie is to be compared to that provided by the Caltech team (\texttt{http://www.its.caltech.edu/~nmvriend/research/multimedia.html}) that never show a sliding technique able to induce an homogeneous steady avalanche. Contrarily to their statements, they do not show either that booming sustains after the avalanche stops (one sees the sliders, not the avalanche). From 88" to 105": comparison of the sliding techniques. [Citation of Auxiliary Material (Animations S1, S2, S3 and S4) provided as scientific complements to their article by Vriend et al.]}
\end{figure}	

 \end{document}